\documentclass[reprint,superscriptaddress,twocolumn,
               amssymb,amsfonts,amsmath]{revtex4-2}
\usepackage[pdftex]{graphicx}
\usepackage[dvipsnames]{xcolor}
\usepackage[pdftex,colorlinks=true,linkcolor=blue,citecolor=blue,urlcolor=blue]{hyperref}
\usepackage{bm}
\usepackage[utf8]{inputenc}
\usepackage[T1]{fontenc}
\setcitestyle{super,sort&compress}

\newcommand{\ilm}{Université Lyon 1, CNRS, LMI, UMR 5615, Villeurbanne, France}
\newcommand{\lmi}{Université Lyon 1, CNRS, LMI, UMR 5615, Villeurbanne, France}
\newcommand{\pprime}{Université de Poitiers, ISAE-ENSMA, CNRS, PPRIME, Poitiers, France}
\newcommand{\umet}{Université Lille, CNRS, INRA, ENSCL, UMET, UMR 8207, Lille, France}

\begin{document}

\title{Deciphering borophene growth pathways with data-driven simulations}

\author{Colin Bousige}
\email{colin.bousige@cnrs.fr}
\affiliation{\lmi}

\author{Jean Furstoss}
\affiliation{\pprime}

\author{Julien Lam}
\affiliation{\umet}

\author{Pierre Mignon}
\affiliation{\ilm}

\date{\today}

\begin{abstract}
    Deterministic synthesis of borophene remains challenging because many polymorphs compete during nucleation and growth.
    Here we combine a reactive machine-learned interatomic potential with grand-canonical Monte Carlo simulations and data-driven structural classification to track borophene formation from early nuclei to extended layers on Ag(111) and Ag(100).
    We build temperature-pressure substrate growth maps and resolve how vacancy motifs, phase intermixing and seed structure govern polymorph selection.
    The simulations reproduce key experimental trends, including the prevalence of $\beta_{12}$/$\chi_3$ phases and their temperature-dependent competition, while revealing kinetic pathways that connect metastable nuclei to long-range order.
    We identify conditions that suppress competing motifs and promote targeted phases, providing actionable synthesis windows.
    These results establish a predictive framework for directing borophene growth and, more broadly, for controlling polymorphism in low-dimensional materials by coupling atomistic simulation with machine-learning-enabled phase recognition.
\end{abstract}

\maketitle

The advent of borophene, a one-atom-thick two-dimensional boron sheet, has marked a milestone in materials science \cite{mannix_synthesis_2015,feng_experimental_2016}.
This material, characterized by its rich polymorphism \cite{tang_novel_2007,mannix_synthesis_2015,feng_experimental_2016,wu_twodimensional_2012,penev_polymorphism_2012}, introduces a metallic 2D system with partial stability.
Its attributes position borophene as a candidate for next-generation technologies, including high-performance, flexible \cite{mortazavi_anomalous_2017,pang_superstretchable_2016,zhang_elasticity_2017}, and transparent \cite{lherbier_electronic_2016} electronics, as well as optoelectronic devices \cite{huang_twodimensional_2017,lian_integrated_2020} and high-capacity ionic batteries \cite{liu_monolayer_2016,jiang_borophene_2016,li_high_2017,liang_borophene_2017,rao_ultrahigh_2017,zhang_borophene_2017}.
Borophene's structural diversity enables precise tuning of its plasmonic, electronic, thermal, and mechanical properties.
A growing body of literature now offers comprehensive reviews on borophene's progress and potential \cite{mannix_borophene_2018,hou_borophene_2020,kaneti_borophene_2021,ou_emergence_2021,kumar_unlocking_2025,wang_borophene_2024,wang_review_2019,adekoya_advances_2024,gupta_borophene_2024,innis_borophene_2025}.

\begin{figure*}[htp]
    \centering
    \includegraphics[width=\linewidth]{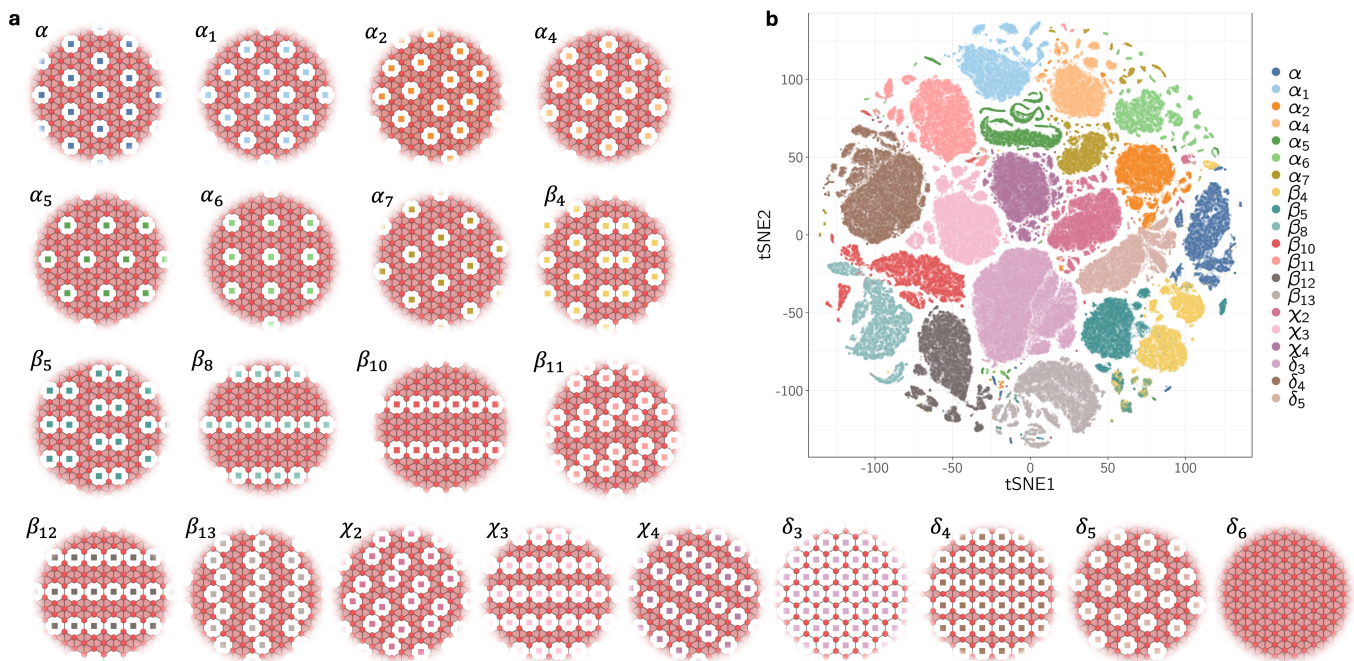}
    \caption{
    \textbf{Allotrope identification.}
    \textbf{a}. Snapshots of the 21 different allotropes included in the training dataset in this study.
    In direct application of our classification model, the colored squares identify the vacancies with their corresponding allotrope. The same colors will be used throughout the article. The nomenclature ($\alpha$, $\beta$, $\chi$, and $\delta$) refers to the coordination of the boron atoms, while the subscript is only used to distinguish between different allotropes (except for the $\delta$ allotropes where it indicates the coordination).
    We selected these allotropes because (1) they include all those experimentally observed on silver \cite{innis_borophene_2025} ($\alpha$, $\beta_{8}$, $\beta_{10}$, $\beta_{12}$, and $\chi_3$) as well as those synthesized on copper, gold, and aluminum ($\alpha_1$, $\alpha_5$, $\chi_3$), and (2) they cover the full range of vacancy densities $\nu$ (from 0 for $\delta_6$ to 1/3 for $\delta_3$), a critical parameter for borophene properties. Some other allotropes are punctually seen during our long growth simulations (e.g. $\alpha_7$), so we also included them in the training dataset to be able to identify them during the growth process.
    \textbf{b}. Dimensionality reduction of the LMBTR descriptors for the different allotropes in the training set through a two-components t-SNE projection. Each point corresponds to a vacancy in the structures obtained during the growth process, and the color corresponds to the allotrope it belongs to. Even with a 2-components t-SNE projection, the different allotropes are quite well separated in the descriptor space. We use a 155-components PCA projection for the actual analysis, which allows for an even better separation.
    }
    \label{fig-some_allotropes}
\end{figure*}

To date, bottom-up synthesis methods have emerged as the most reliable approach for producing monolayer borophene \cite{innis_borophene_2025}.
Experimental efforts have identified eleven distinct polymorphs, each influenced by synthesis parameters such as temperature and substrate type and orientation \cite{mannix_synthesis_2015,feng_experimental_2016,li_experimental_2018,vinogradov_singlephase_2019,zhong_metastable_2017,zhong_synthesis_2017,kiraly_borophene_2019,mazaheri_chemical_2021,wu_largearea_2019,wu_micrometrescale_2022,tai_synthesis_2015,sutter_largescale_2021,radatovic_macroscopic_2022,omambac_segregationenhanced_2021,cuxart_borophenes_2021,sheng_raman_2019,wu_synthesis_2025,innis_borophene_2025}.
To systematically investigate these polymorphs, we compiled a dataset of 21 distinct borophene allotropes (including all those that have been observed experimentally, as well as others), as shown in Fig.~\ref{fig-some_allotropes}.
In addition, phase intermixing is frequently observed \cite{liu_intermixing_2018,li_epitaxial_2023,kiraly_borophene_2019,wang_realization_2020,wu_synthesis_2025}, underscoring the complexity of the borophene system.
Yet, this very complexity also hints at the possibility of selectively synthesizing specific polymorphs tailored to desired properties -- provided the optimal experimental conditions are identified.
Consequently, unraveling the growth mechanisms of borophene has become a critical priority, as achieving deterministic synthesis is essential to harnessing its full potential across a spectrum of applications.

Experimentally, there have been no systematic studies of borophene growth allowing identifying early steps in the expansion mechanism or possible precursors leading to particular allotropes -- or to phase mixing of specific allotropes.
This is further complicated by the fact that the borophene growth mechanism is substrate-dependent: it was shown that the dose of atomic boron required to observe nanoscale borophene islands on Au(111) is an order of magnitude greater than that required to form a monolayer on Ag(111) \cite{kiraly_borophene_2019}.
It was interpreted as an indication that boron dissolves into the bulk in Au(111) -- leading to a surface segregation growth mechanism -- while remaining on the surface in Ag(111).
This is consistent with the lower enthalpy of formation of gold borides compared to silver borides, although it is positive in both cases \cite{niessen_enthalpy_1981}.
Although gas-phase photoelectron spectroscopy combined with theoretical atomistic calculations predicted several planar structures stabilized by the presence of hexagonal vacancies for B$_{30}$, B$_{36}$, B$_{37}^-$ or B$_{38}^-$ clusters \cite{sergeeva_understanding_2014, li_epitaxial_2023, kiran_planartotubular_2005, li_b30_2014, chen_planar_2017, piazza_planar_2014}, there is no experimental evidence for their existence on metal substrates.

Only a few studies have investigated the growth and nucleation of borophene on metal substrates \cite{liu_boron_2013, xu_nucleation_2016,shu_unveiling_2016,yu_mechanism_2026}, and most have focused on the very early stages of growth.
It was shown from both ab initio molecular dynamics (MD) \cite{shu_unveiling_2016} and static DFT calculations \cite{xu_nucleation_2016} that the presence of hexagonal vacancies stabilizes an initial triangular borophene cluster and that early borophene structures are embedded in the first metal layer.
Very recently, a study using a machine learned interaction potential (MLIP) and metadynamics \cite{yu_mechanism_2026} has focused on the phase transition from the $\beta_{12}$ to the $\chi_3$ allotrope in extended surfaces, highlighting a mechanism involving boron dissolution into the silver substrate.

Here, we address this challenge by combining a reactive MLIP \cite{mignon_neural_2023,bousige_portable_2025} with grand-canonical Monte Carlo (GCMC) simulations and data-driven structural classification to track borophene formation from early nuclei to extended layers on Ag(111) and Ag(100).
We identify conditions that suppress competing motifs and promote targeted phases, providing actionable synthesis windows.
These results establish a predictive framework for directing borophene growth and, more broadly, for controlling polymorphism in low-dimensional materials.

\section*{Data-driven classification of borophene allotropes}

To process the millions of snapshots generated in this study, we designed a fast algorithm to locate vacancies and a machine-learning strategy to classify them based on local structural descriptors.
It is indeed the organization of vacancies rather than the B-B pair distribution function that defines the allotropes (Fig.~S2).

After testing several descriptors, we retained the LMBTR one for its superior phase separation performance.
A training dataset comprising the 21 allotropes (Fig.~\ref{fig-some_allotropes}\textbf{a}) was generated, and dimensionality reduction was applied using PCA (99.9\% retained variance).
A two-component t-SNE projection (Fig.~\ref{fig-some_allotropes}\textbf{b}) demonstrates excellent separation of most allotropes in descriptor space, while a 155-component PCA projection was used for the actual analysis.

A Random Forest classifier was trained on an 80:20 split of the dataset, achieving an accuracy of 1.0 on the training set and 0.9978 on the testing set.
An example of the classification is shown in Fig.~S6, where the different allotropes in a structure from a large-scale GCMC growth simulation at 700~K are identified.
Vacancies classified as ``unknown'' are primarily located at island edges, where local environments deviate most from the training set.
This approach enables robust quantification of allotrope populations as a function of simulation conditions.

\section*{Full-scale growth of borophene on silver}

\begin{figure*}[htp]
    \centering
    \includegraphics[width=.9\linewidth]{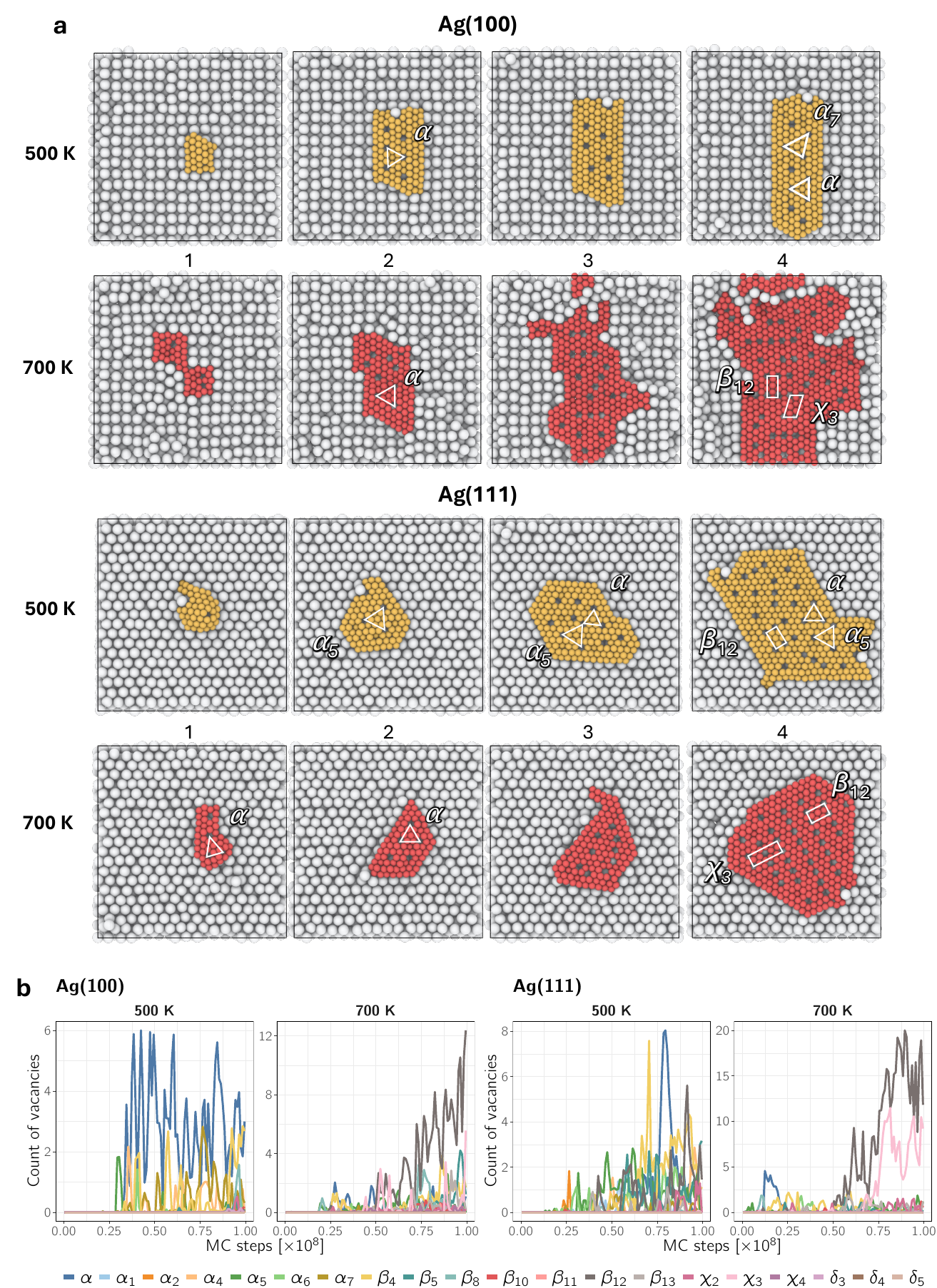}
    \caption{\textbf{Analysis of a full-scale growth}.
        \textbf{a.} Snapshots of the structures obtained at regular stages of the growth process on Ag(111) and Ag(100).
        The box size is $\sim50\times50$~\AA$^2$, and some identifiables allotropes are highlighted in white.
        \textbf{b.} Number of vacancies of each allotrope along the full-scale GCMC growth process. The unknown (disordered) vacancies are not shown for clarity.}
    \label{fig-full_growth}
\end{figure*}

First, to capture the full nucleation-and-growth sequence, we have simulated the full-scale growth of borophene on Ag(111) and Ag(100) at 500~K and 700~K.
We start from the bare silver surface with an initial seed of 7 boron atoms forming an hexagonal cluster placed at the surface's center.
To improve statistics, 5 simulations were run in parallel with different random seeds.
Nucleation being a stochastic process, the growth of borophene is not guaranteed to happen in a given simulation, and the growth rate is strongly dependent on the pressure and temperature.
Moreover, GCMC is a slow method because it is not fully parallelizable -- especially in the case of growth studies where rare events have to be observed for the growth to happen -- so pressures higher than the ones seen in ultra-high vacuum have to be used to accelerate the growth process.
As a compromise, to be as close as possible to the experimental conditions while maintaining a realistic time frame for the simulations, a pressure of $10^{-3}$~Pa at 500~K and $1$~Pa at 700~K were applied.
Of the 5 simulations run in parallel, only the ones successful in nucleating are shown here -- a testimony of the stochastic nature of the process.
The simulations were stopped before the growing islands could occupy the whole surface in order to avoid edge effects, which happened after a run of 10 to 50~ns -- corresponding to 1 to $5\times10^8$ MC steps along the trajectory.

Interestingly, growth paths at both temperatures seem to go through similar early stages, with the formation of the allotrope $\alpha$ or $\alpha_5$ when the island has a symmetric shape adapted to the substrate (Fig.~\ref{fig-full_growth}\textbf{a}).
As the island grows larger, more disorder appears, which further evolves into phase mixing.
At later stages, the growth at 700~K favors the formation of $\beta_{12}$ and $\chi_3$, while the growth at 500~K favors the formation of hexagonal allotropes such as $\alpha$, $\alpha_5$, and $\beta_{12}$ -- the $\alpha$ domain being the largest in the 500~K case.
Moreover, for Ag(111), we observe that the cluster edges follow geometric forms due to the arrangement of the substrate surface atoms, i.e., epitaxial growth.
At high temperature, we also see in Fig.~\ref{fig-full_growth}\textbf{a} that the $\beta_{12}$ and $\chi_3$ vacancies tend to follow the surface orientation of the Ag atoms.

Figure~\ref{fig-full_growth}\textbf{b} shows the evolution of the different allotropes proportion during all the GCMC growth processes from Fig.~\ref{fig-full_growth}\textbf{a}.
We see that the highest temperature of 700~K always favors the intermixed growth of $\beta_{12}$ and $\chi_3$, while the lower temperature of 500~K goes through a more complex path with many different intermediate phases, but with a predominance of $\alpha$.
This is in agreement with the experimental observations of borophene growth on Ag(111) at 570~K \cite{zhong_metastable_2017} and above 700~K \cite{innis_borophene_2025,campbell_resolving_2018,zhong_metastable_2017,sheng_raman_2019}, where mixed phases of $\beta_{12}$ and $\chi_3$ as well as $\alpha$ are observed at low temperature synthesis conditions, while at higher temperatures only the $\beta_{12}$ and $\chi_3$ phases are observed as single domains.

While these simulations highlight the nucleation and growth processes and confirm experimental observations, they do not fully disentangle the contributions of nucleation and growth to the final polymorph distribution.
A more sophisticated approach is necessary to understand the growth process in more detail.

\section*{Separating growth from nucleation}

A central difficulty in understanding borophene formation is that nucleation and growth are entangled processes: the allotrope that eventually covers the surface depends both on which nucleus forms first and on how that nucleus subsequently expands.
Yet, nucleation is notoriously difficult to simulate: it is a rare, stochastic event that requires prohibitively long simulation times to observe spontaneously, making it impractical to study systematically -- and with meaningful statistics -- across the full parameter space of temperatures, pressures, and substrate orientations.
To disentangle these contributions and also establish the thermodynamic baseline against which kinetic effects can be assessed, we designed a two-tier simulation strategy.

First, to establish the intrinsic thermodynamic stability of each polymorph on the substrate, we performed MD melting simulations: pre-formed islands of each allotrope are heated stepwise until the phase loses structural coherence, yielding a substrate-dependent melting temperature $T_m$ for all 21 allotropes.
This allows us to rank the relative stability of all allotropes on both Ag(111) and Ag(100), and to identify which phases are stable enough to survive the synthesis temperatures used experimentally.

Second, to probe growth independently of nucleation, we ran short GCMC simulations starting from the same pre-formed seeds (Fig.~\ref{fig-some_allotropes}) at controlled temperature and chemical potential.
By imposing the initial polymorph, we bypass the nucleation step entirely and directly measure how favorably each phase recruits additional boron atoms from the vapor -- and whether it retains its identity or transforms into a competing phase.

In order to provide statistics on the results, simulations in both tiers are performed for each allotrope on each surface (Ag(100) or Ag(111)), with different initial seed shapes (circular or square, Fig.~S7), using 11 different random rotations and starting atomic velocities.
Together, these two tiers allow separating the effects of thermodynamic stability and self-propagation ability during the growth process to identify whether the nucleation pathway biases the final polymorph distribution.

\section*{Analysis of the stability of the different allotropes: melting temperatures}

\begin{figure}[htp]
    \centering
    \includegraphics[width=\linewidth]{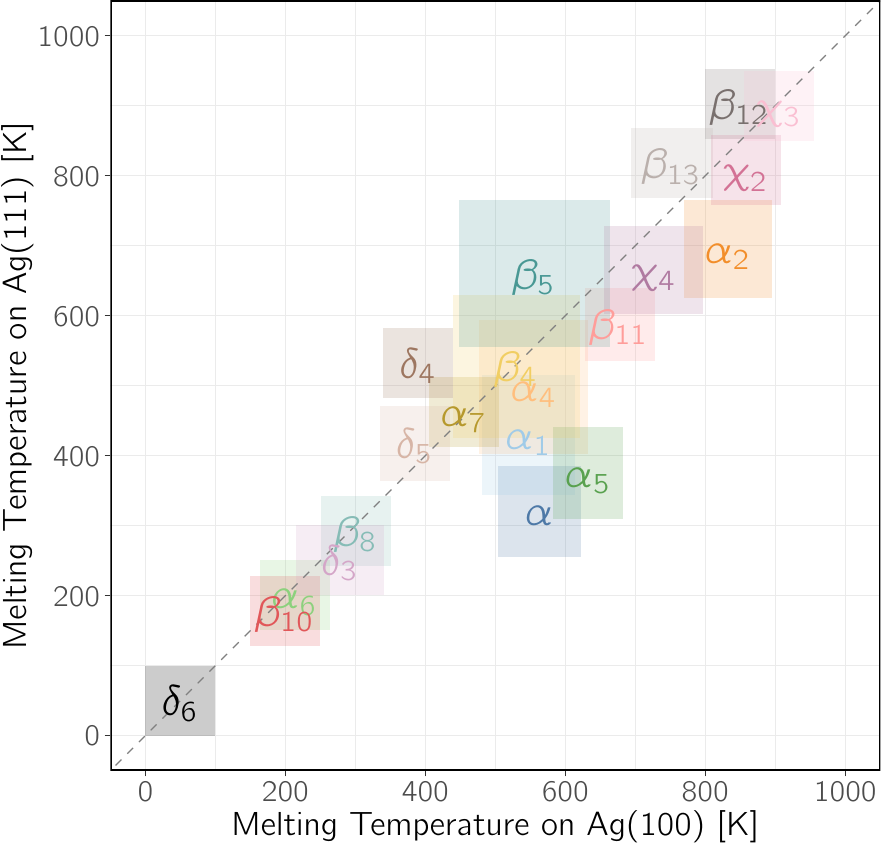}
    \caption{\textbf{Melting temperatures for the different allotropes on Ag(111) versus the one on Ag(100)}. The rectangles are centered on the average melting temperature for each allotrope and each surface, and their width and height are given by the standard errors $SE$ of the melting temperatures obtained for the $N=22$ different simulations for each allotrope (11 random orientations, 2 shapes), with $SE=\max\left(d_{T_m}, \sigma_{T_m}/\sqrt{N}\right)$.
    Here, $d_{T_m}= 50$~K is the error on the determination of each melting temperature $T_m$, and $\sigma_{T_m}$ the standard deviation of $T_m$. The dashed line corresponds to the case where the melting temperature is the same on both surfaces.}
    \label{fig-melts}
\end{figure}

The average melting temperatures obtained for each allotrope on Ag(111) and Ag(100) is shown in Fig.~\ref{fig-melts}.

Unsurprisingly, the $\delta_6$ allotrope is the least stable, with a melting temperature of around 50~K -- it is the only one that does not have any vacancy in its structure, and thus the only one that does not benefit from the stabilization provided by their presence.
Mannix \emph{et al.} have claimed to have synthesized $\delta_6$ on Ag(111) at 550\,$^\circ$C in their seminal paper \cite{mannix_synthesis_2015}, but it was later refuted by Feng \emph{et al.} \cite{feng_experimental_2016} who proposed $\beta_{12}$ instead.
Our results confirm that it should indeed be very difficult to synthesize $\delta_6$ on silver.

The most stable allotropes appear to be $\chi_3$, $\beta_{12}$, $\chi_2$, $\beta_{13}$, $\alpha_2$, $\chi_4$, in decreasing order of stability.
Experimentally, $\beta_{12}$ and $\chi_3$ are indeed the most commonly observed allotropes on Ag(111), with a $\beta_{12}$ predominance below 850~K, a coexistence between 850~K and 950~K, and a $\chi_3$ predominance above 950~K \cite{innis_borophene_2025,yu_mechanism_2026}.
The allotropes $\alpha_2$, $\chi_4$, $\beta_{13}$ and $\chi_2$ have however never been identified to the best of our knowledge, and $\alpha$ was only observed at lower synthesis temperature (570~K) on Ag(111) \cite{zhong_metastable_2017}.
On Ag(110), the allotropes $\chi_3$, $\beta_{12}$, $\beta_{8}$ and $\beta_{10}$ were observed from synthesis at 570~K \cite{zhong_synthesis_2017}, but from our results, $\beta_{8}$ and $\beta_{10}$ seem to be in the group of the least stable allotropes on Ag(100) and Ag(111).
There is only one reported synthesis of borophene on Ag(100), and it shows phase intermixing between the $\chi_3$ and $\beta_{12}$ allotropes \cite{wang_realization_2020}.
This was interpreted as a substrate-commensuration effect: intermixing would be possible on Ag(100), but not on Ag(111), where both allotropes are incommensurate with the substrate \cite{li_epitaxial_2023}.
However, intermixing was also reported on Ag(111) \cite{liu_intermixing_2018,gao_substrateinduced_2026}, which argues against a purely commensuration-based explanation.
A more consistent interpretation would be that intermixing is promoted by lower synthesis temperatures in both cases ($\sim720$~K on Ag(111) and $\sim550$~K on Ag(100)), where neither allotrope is strongly stabilized over the other.
This is supported by a recent metadynamics study on Ag(111): $\chi_3$ becomes more stable than $\beta_{12}$ only above about 950~K, whereas $\beta_{12}$ is favored at lower temperature \cite{yu_mechanism_2026} -- as is observed experimentally \cite{liu_intermixing_2018}.

Finally, while most allotropes are equally stable on Ag(111) and Ag(100) (Fig.~\ref{fig-melts}(b)), some are more stable on one surface than on the other.
This is the case for $\alpha$ and $\alpha_5$, which are stable at higher temperatures on Ag(100) compared to Ag(111) by a noticeable margin.

To understand the fact that stable allotropes are not observed experimentally, a deeper look at the growth process is now necessary.

\section*{Analysis of the growth of borophene from various starting configurations}

An initial seeded GCMC growth was performed at 3 pressures ($10^{-3}$, $10^{-2}$, and $10^{-1}$~Pa) and 5 temperatures (500, 600, 700, 800, and 900~K) for each allotrope on both Ag(111) and Ag(100) (Figs.~S9-S10).
Concentrating on the lowest pressure in this representative set of simulations, growth on Ag(111) is strongly temperature dependent: at 500~K only $\alpha$ and $\alpha_4$ self-propagate, at 600~K this behavior extends to $\beta_{5}$, $\beta_{12}$, $\chi_3$, and $\alpha_1$, and at 700--800~K it narrows mainly to $\beta_{12}$ and $\chi_3$ (with $\chi_2$ also favorable at 800~K), whereas at 900~K no phase remains favorable and the initial seed progressively disappears at all studied pressures -- the latter temperature was thus omitted in further simulations.
Ag(100) shows the same overall trend, with $\alpha_4$ and $\alpha_7$ favored at 500~K and $\beta_{12}$ and $\chi_3$ favored at 700~K.
While it looks like increasing pressure might allow promoting other allotropes at intermediate temperatures, the effect is not systematic and the overall trend remains the same.
In the following, statistics over 10 new random orientations were acquired only for $P=10^{-3}$~Pa to be closer to experimental high to ultra-high vacuum conditions while keeping the simulation time tractable.

\begin{figure}[htp]
    \centering
    \includegraphics[width=\linewidth]{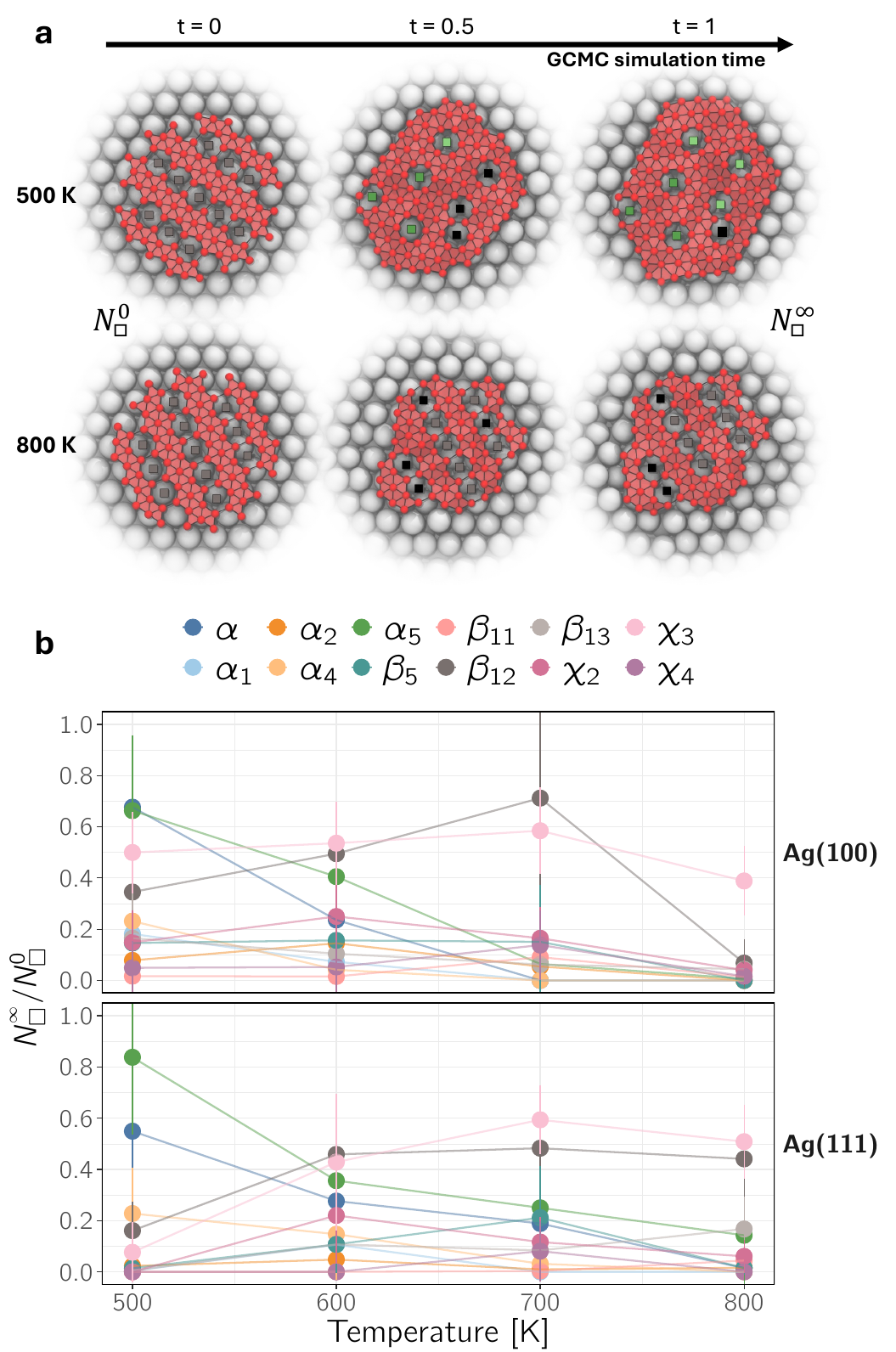}
    \caption{\textbf{Growth of borophene on silver from various starting configurations}.
        \textbf{a.} Illustration of the GCMC self-propagation process for the $\beta_{12}$ allotrope on Ag(111) at 500~K and 800~K. The $N_{\square}^0$ and  $N_{\square}^{\infty}$ quantities are the number of vacancies attributed to the $\beta_{12}$ allotrope at the beginning and at the end of the simulation, respectively.
        \textbf{b.} Relative variation of the number of vacancies attributed to the allotrope that was used in the seed during the growth process as a function of the temperature and orientation. The variation is taken as the last value of the simulation from Figs.~S9-S10 divided by the first one, $N_{\square}^\infty / N_{\square}^0$. Statistics were acquired only for $P=10^{-3}$~Pa, which is the pressure used in this plot. The plotted value is averaged over the 22 different simulations for each allotrope (11 random orientations, 2 shapes). The error bars are given by the standard deviation. Only the allotropes stable above 500~K are shown here (as determined from Fig.~\ref{fig-melts}).}
    \label{fig-seed_growth}
\end{figure}

In Fig.~\ref{fig-seed_growth}, we present the self-propagation metric, $N_{\square}^\infty / N_{\square}^0$, averaged over all 22 simulations for each allotrope stable above 500~K (Fig.~\ref{fig-melts}).
Here, $N_{\square}^0$ is the number of vacancies attributed to the seed allotrope at the start of the simulation, while $N_{\square}^\infty$ is the number of vacancies attributed to this allotrope at the end ($t=0$ and $t=1$ in Fig.~\ref{fig-seed_growth}\textbf{a}, respectively).
We observe on Fig.~\ref{fig-seed_growth}\textbf{a} that edges rapidly rearrange into stable configurations, regardless of the allotrope or seed shape.
Consequently, all values in Fig.~\ref{fig-seed_growth}\textbf{b} are below 1, indicating that seeds rearrange at their edges during early growth.
This rearrangement reduces the number of vacancies attributed to the initial seed allotrope, as edges are inherently unstable and adapt dynamically to minimize energy.
Instead, the atomic arrangement of a given allotrope emerges in the seed's interior, where vacancies form via atomic migration to stabilize the structure.
This behavior is consistent across all seeds, orientations, and initial geometries, as evidenced by the rapid decrease in the $N_{\square}^\infty / N_{\square}^0$ metric with simulation time (Figs.~S9-S10) and our methodology, which combines high-temperature GCMC exploration with short MD relaxations to sample equilibrium structures.

Overall, for $\alpha$ and $\alpha_5$ a good growth propagation is observed at 500~K, then it deceases down to disappearance with increasing temperature on both surfaces.
$\beta_{12}$ and $\chi_3$ are the only consistently self-propagating phases at $\geq700$~K, with increasing $\chi_3$ dominance at higher temperature ($\beta_{12}$ completely disappears at 800~K on Ag(100)).
We note that, apart from $\beta_{12}$ and $\chi_3$, most phases that were observed to be thermodynamically stable remain kinetically disfavored.

On Ag(111), the experimentally-observed competition between $\beta_{12}$ and $\chi_3$ as a function of temperature\cite{liu_intermixing_2018} seems reproduced in our simulations, with a higher retention of $\beta_{12}$ at lower temperature and a higher retention of $\chi_3$ at higher temperature.
We could however not reproduce the predominance of $\chi_3$ above 800~K since at these (T,P) conditions with our GCMC simulations the initial seed completely evaporates.

These trends confirm that propagation kinetics, rather than thermodynamic stability alone, governs polymorph selection under growth conditions.
In other words, polymorph selection follows a two-step filter in our simulations: phases must first be thermodynamically stable enough to survive under synthesis conditions, and then kinetically able to self-propagate during growth.
Several allotropes satisfy the first criterion but fail the second, as seen from their low $N_{\square}^{\infty}/N_{\square}^{0}$ values and frequent phase conversion.
By contrast, $\alpha$ and $\alpha_5$ (lower temperatures), and $\beta_{12}$ and $\chi_3$ (higher temperatures) are the only allotropes that consistently pass both filters on Ag(111) and Ag(100), which explains their experimental predominance.

Overall, our seeded-growth simulations provide a mechanistic explanation for the experimentally observed polymorph selection: $\alpha$ (low T), $\beta_{12}$ and $\chi_3$ (higher T) are not only thermodynamically stable, but also kinetically favored to propagate during growth, while other allotropes either fail to nucleate or convert into more favorable phases.

\section*{Discussion and Outlook}

The central message of this work is that borophene polymorph selection on silver cannot be predicted from stability alone.
Several allotropes rank among the most stable in our melt analysis (for example $\beta_{13}$, $\chi_{4}$, $\chi_{2}$ and $\alpha_{2}$), yet they do not efficiently propagate during growth.
Conversely, $\beta_{12}$ and $\chi_3$ combine high stability with favorable growth kinetics, which naturally explains why they dominate experimental reports.
Mechanistically, our seeded-growth data suggest that $\beta_{12}$ and $\chi_3$ exhibit more permissive propagation fronts.
Unlike other allotropes, their growth is not governed by edge rearrangements that constantly adapt to incoming atoms.
Instead, vacancies diffuse within the interior of the boron cluster, driving the system toward the most stable atomic arrangement.
This explains their robustness under a wide range of synthesis conditions.
Their vacancy-rich motifs also appear to better accommodate substrate registry on both Ag(111) and Ag(100), making boundary disorder easier to heal into continued $\beta_{12}$/$\chi_3$ growth.
This thermodynamics-versus-kinetics decoupling provides a consistent framework to rationalize the gap between calculated relative stabilities and observed phase prevalence.

Our simulations reveal that temperature acts as a discriminative kinetic selector in borophene growth.
At lower temperatures, growth proceeds through a broader set of intermediates, favoring hexagonal motifs such as large $\alpha$ domains.
In contrast, higher temperatures drive the robust development of $\beta_{12}$ and $\chi_3$ on both Ag(111) and Ag(100).
These trends align with experimental observations of $\alpha$-rich or mixed phases near 570~K \cite{zhong_metastable_2017} and predominantly $\beta_{12}$/$\chi_3$ phases at higher synthesis temperatures \cite{innis_borophene_2025,campbell_resolving_2018,zhong_metastable_2017,sheng_raman_2019,liu_intermixing_2018}.
Practically, our maps indicate that high-temperature growth is the most reliable route toward phase-pure films, while lower-temperature conditions are better suited for targeting metastable or mixed polymorph landscapes.

Beyond borophene, the methodology introduced here -- combining controlled seeding, long-time GCMC growth, and local machine-learning phase recognition -- is directly transferable to other polymorphic low-dimensional materials where nucleation and growth are similarly entangled.
However, extending this framework to larger lateral scales, longer times, and richer experimental environments (e.g., explicit time-dependent deposition fluxes, step edges, and defects) will require substantial computational resources.
Coupling the simulations to in situ observables for direct, time-resolved validation would further enhance their predictive power.
Such developments, while computationally demanding, are essential for enabling kinetics-aware synthesis design rather than post hoc phase assignment.

\section*{Acknowledgements}

C.B. acknowledges support by the French National Research Agency grant (ANR-21-CE09-0001-01).
This work was granted access to the HPC resources by GENCI: to IDRIS under the allocation 2022-(Grant A0120807662), and to CINES under the allocation AD010915969R1.

\section*{Data Availability Statement}
The data and code that support the findings of this study are openly available in on GitHub and Zenodo \cite{colinbousige_boroml_2026}.

\section*{Competing financial interests}
The authors declare no competing financial interests.

\section*{Author's contributions}

C.B. acquired funding, designed and carried out all steps of the study, and wrote the manuscript.
P.M., J.F. and J.L. contributed to the analysis of the results.
All authors discussed and revised the manuscript.

\section*{Methods}

\subsection*{MLIP architecture}

We used the high-dimensional feed-forward neural network potential developed by Behler and Parrinello \cite{behler_generalized_2007} and implemented in the n2p2 v2.3.0 software \cite{singraber_parallel_2019,singraber_librarybased_2019,singraber_compphysvienna_2021}.

In this method, the input layer corresponds to the geometric descriptors of the system, treated by hidden layers of a neural networks (usually two) made up of a defined number of neurons each.
A neural network is defined for each element of the system, resulting in atomic energies and forces (output layers).
Atomic environments, defined around each atom $i$ by shells of radius $r_c$ (cutoff radius), are then described by a vector of radial and angular symmetry functions, $G_i$, which describe the local environment of each atom in the system in terms of 2- and 3-body densities \cite{behler_atomcentered_2011}.
The ensemble of $G$ functions forms the input layer of the NNP.

Here we used a set of 22 radial and 30 angular symmetry functions per element, resulting in an input dimension of 104 for the neural network.
The cutoff radius was set to 6.35~\AA: it is large enough to include all atoms in the first coordination sphere of each atom, but small enough to keep the computational cost reasonable.
All parameters of the symmetry functions are provided on Zenodo \cite{colinbousige_boroml_2026} -- they are the same ones that were used in our previous work \cite{bousige_portable_2025}.
Thorough validation of the MLIP was performed in our previous works \cite{mignon_neural_2023,bousige_portable_2025}.

\subsection*{Molecular Dynamics simulations}

All MD simulations are run with LAMMPS (v. 22Jul2025) \cite{plimpton_fast_1995,thompson_lammps_2022}, together with the ML-HDNNP package implementing n2p2 \cite{behler_generalized_2007,singraber_parallel_2019,singraber_librarybased_2019} (v. 2.3).
For these simulations, the bottom layer of the metal substrate is kept fixed, and all atoms above are allowed to move.
The mobile atoms are thermalized in the NVT ensemble using a Nosé-Hoover thermostat with time steps of 1~fs and a coupling time of 100~fs.

To determine the melting temperature, we use a ladder procedure alternating 1~ns plateaus at a given temperature, separated by 0.1~ns temperature ramps.
The plateaus temperatures start at 10~K, followed by a step at 50~K, and then every 50~K up to 1,200~K.
Frames are saved every 10~ps during the plateaus, resulting in 101 frames per trajectory.
For this procedure, we use $\sim35\times35$~\AA$^2$ Ag(100) or Ag(111) slabs with borophene seeds of size 20~\AA\ (ca. 100 atoms, depending on the allotrope).
These seeds are either circular or square, and 11 simulations where the borophene seeds are randomly rotated on the Ag(100) or Ag(111) surface are run in parallel.
There are thus 2 Ag orientations $\times$ 2 shapes $\times$ 21 allotropes $\times$ 25 temperatures $\times$ 11 random orientations $=$ 23,100 trajectories to analyze, with 101 frames each.
To analyze the melts, one needs a definition of a melting threshold.
We arbitrarily define it like this: for each starting allotrope, we record the number of vacancies in the initial configuration.
Then, we define $T_{\text{m}}$ as the temperature at which at least 10\,\% of the trajectory contains configurations with fewer than 50\,\% of this initial number of vacancies classified as belonging to the starting allotrope.
The results of the classification of the different allotropes during a melt of $\alpha$ on Ag(111) are shown in Fig.~S8 as an example.
Looking at the dark blue line, which corresponds to the $\alpha$ allotrope, we can see that it starts to decrease significantly and irremediably at 400~K, as the number of vacancies attributed to $\alpha$ starts to decrease and the number of unattributed vacancies starts to increase: for this configuration, 400~K is thus defined as the melting temperature $T_m$.

For generating data to train the classification algorithm, we use large circular islands of borophene (diameter 85~\AA\ on a 100~\AA\ silver substrate) to avoid edge effects and make the clusters' structures stable (larger clusters are less prone to structural changes in a short simulation time).
In these simulations, we ramp the temperature from 10~K up to 700~K in 0.2~ns to introduce thermal fluctuations.
If a given allotrope is not stable above a given temperature, the simulation is stopped at this temperature and all the previous structures are included in the dataset.

\subsection*{GCMC growth simulations}

GCMC simulations were run using the MC package of LAMMPS.
For these simulations, the bottom layer of the metal substrate is kept fixed, and all atoms above are allowed to move.
In these GCMC simulations, we alternate between MD steps (where all atoms are allowed to move) and MC steps (where only boron atoms are inserted or deleted and no displacement is allowed).
During MD steps, the mobile atoms are thermalized in the NVT ensemble using a Nosé-Hoover thermostat with time steps of 1~fs for a relaxation time of 1~ps.
After each thermalization block, $10^4$ MC insertion and deletion attempts are performed for boron atoms, and attempts are accepted or not based on the Metropolis criterion \cite{frenkel_understanding_2002}.
To accelerate the growth process, boron atoms are inserted at random positions directly on the surface, i.e. $+4$~\AA\ and $-1$~\AA\ around the $z$ position of the top silver layer, thus allowing for multilayered borophene growth as well as solubilization of boron atoms within the silver lattice.

We ran two types of simulations: long ones, and short ones.
In all cases, the wanted pressure and temperature are imposed by defining the chemical potential:
\begin{equation}
    \begin{aligned}
        \mu(T,P) & =\mu^{trans}+\mu^{ex}                                    \\
                 & =k_BT\ln\left(\frac{P\Lambda^3}{k_BT}\right)+\mu^{ex}(T)
    \end{aligned}
\end{equation}\label{eq-mu}%
where $\mu^{trans}$ is the chemical potential coming from the atomic translational degrees of freedom, $\Lambda = h/\sqrt{2\pi m_Bk_BT}$ is the thermal de Broglie wavelength, $m_B$ the mass of a boron atom, $k_B$ the Boltzmann constant, and $\mu^{ex}$ is the excess chemical potential due to energetic interactions.
The latter is determined using the Widom insertion method.
Full details are provided in Supplementary Note 1 together with a discussion on the evolution of $\mu^{ex}$ as a function of temperature and substrate orientation.

For the short simulations (Fig.~\ref{fig-seed_growth}), we use $\sim35\times35$~\AA$^2$ Ag(100) or Ag(111) slabs with borophene seeds of size 20~\AA\ (ca. 100 atoms, depending on the allotrope).
These seeds are either circular or square, and are randomly rotated on the Ag surface.
Eleven simulations are run in parallel for $2\times10^5$ time steps.
Each take about 2h to complete, and there are 2 Ag orientations $\times$ 2 shapes $\times$ 21 allotropes $\times$ 5 temperatures $\times$ 11 random orientations $=$ 4620 simulations to run.
In the long simulations (Fig.~\ref{fig-full_growth}), we use $\sim50\times50$~\AA$^2$ Ag slabs with an hexagonal cluster of 7 boron atoms in the center of the surface as starting configuration.
We let the system evolve for at least $10^7$ time steps ($10^8$ MC attempts) and up to $5\times10^7$ time steps -- we stop the simulations before full coverage is reached to avoid edge effects. We run 5 simulations in parallel with different random seeds. These simulations typically take around 20 days for $10^7$ time steps on a 128 cores node.

\subsection*{Vacancies detection}

To analyze over three million snapshots efficiently, we require a fast and automated method for precisely locating these vacancies.
A complex algorithm based on atom positions and coordination would be computationally expensive and impractical for this scale.
Therefore, we adopt a straightforward yet effective approach by treating the problem as an image processing task.

In our method, boron atoms are represented as disks, and a binarized image of the boron structure projected on $(x,y)$ plane is generated.
Using the \texttt{OpenCV} Python library \cite{bradski_opencv_2026}, we apply blob detection to identify vacancy locations in the $(x,y)$ plane, as illustrated in Fig.~S3.
The $z$ coordinate of each vacancy is set to the average $z$ position of all boron atoms in the snapshot.
This approach is highly efficient, and the process can be easily parallelized on a list of snapshots, making it feasible to analyze millions of structures in a timely manner.

\subsection*{Allotrope classification}

To classify the different allotropes, we use a machine learning strategy based on local atomic descriptors, such as several structural analysis studies \cite{becker_unsupervised_2022,furstoss_allaround_2025,goryaeva_reinforcing_2020}.
The local atomic descriptor we ended up using is defined as the concatenation of the radial and angular LMBTR descriptor \cite{huo_unified_2022} as provided by the \texttt{DScribe} Python library \cite{himanen_dscribe_2020}.
We used a cutoff radius of 15~\AA, a grid spacing of 0.5~\AA\ for the radial part and 1$^\circ$ for the angular part, and a $w_{scale}=0.4$ parameter for the weighting function.
This descriptor was chosen after testing several other descriptors, including ACSF\cite{behler_generalized_2007,behler_atomcentered_2011}, SOAP \cite{bartok_representing_2013}, bond-orientational order \cite{steinhardt_bondorientational_1983}, or ACE \cite{drautz_atomic_2019,lysogorskiy_performant_2021}.
While most descriptors performed well on the training set with large islets, LMBTR was the only one that allowed for proper classification when applied to smaller islets as used in the study.
With the LMBTR descriptor, we have a 600-dimensional vector for each vacancy.
We then use a PCA dimensionality reduction technique with a 99.9\,\% variance threshold, reducing the dimensionality of the descriptor space down to 155.

To generate the training data for the classifier, we use large circular islands of borophene (diameter 85~\AA\ on a 100~\AA\ silver substrate) to avoid edge effects and make the clusters' structures stable.
If a given allotrope is not stable above a given temperature, the simulation is stopped at this temperature and all the previous structures are included in the dataset.
The case of the $\delta_6$ allotrope is particular as it has no vacancies, so it is not included in the training set and cannot be identified in the structures obtained during the growth process.
But all our simulations show that $\delta_6$ is unstable at all temperatures, so it is not expected to be present in the structures obtained during the growth process.
After splitting the dataset into training and testing sets (80:20), we use a Random Forest (100 trees, default parameters) to classify the different allotropes based on their reduced-LMBTR descriptors.
A threshold of 0.4 is used for identification, meaning that if the probability of belonging to a given allotrope is higher than 0.4, the vacancy is classified as belonging to this allotrope, otherwise it is classified as ``unknown''.


\providecommand{\noopsort}[1]{}

\end{document}